# CURRENT AND PLANNED HIGH PROTON FLUX OPERATIONS AT THE FNAL BOOSTER*

F.G. Garcia[#], W. Pellico, FNAL, Batavia, IL 60510, USA

*Abstract*

The Fermilab Proton Source machines, constituted by Pre-Injector, conventional Linac and Booster synchrotron, at Fermi National Accelerator Laboratory (Fermilab) had have a long history of successful beam operations. Built in late 60's, the Fermilab Proton Source began operations early in the 70's and since then it has successful provided protons to support the laboratory physics experiments. During the past decade, Booster performance reached unprecedented proton flux delivery of the order of 1.0–1.1E17 protons per hour, corresponding to 40 kW of beam power while maintained an allowed upper limit of 525 W of beam loss in the tunnel. In order to achieve this historical performance, major hardware upgrades were made in the machine combined with improvements in beam orbit control and operational awareness. Once again, the Proton Source has been charged to double their beam throughput, while maintaining the present residual activation levels, to meet the laboratory Intensity Frontier program goals until new machines are built and operational to replace them. In this paper we will discuss the plans involved in reaching even higher beam throughput in Booster.

## INTRODUCTION

The Fermilab Booster [1] is rapid cycling synchrotron which accelerates protons from 400 MeV to 8 GeV for injection, until recently exclusively into the Main Injector, and for the Booster Neutrino beamline (BNB). It is 472 m in circumference and it operates at a Radio Frequency (RF) harmonic number of 84. The Booster frequency changes rapidly through the accelerator cycle, from 37.9 – 52.8 MHz and there are 19 RF cavities in the machine. Early in 2000's the demand for protons increased 12-fold in comparison to the previous 10 years of Booster operations prompted by the beginning of the neutrino program at Fermilab.

Present protons per batch in Booster are 4.5E12 at 7.5 Hz with 90% efficiency and 85% uptime. In the future, the required number of protons to support the laboratory physics program, until Project X is operational, requires double the proton flux from the present running conditions. Booster is not current capable of running at 15 Hz mainly due to RF power system limitations [2]. In addition, reliability of Booster machine is an issue that has been increasing and will continue to increase with time just as the physics program demands better performance. Improvements in both hardware and operational efficiency of the Booster are required in order to have a successful physics program.

Therefore, the Proton Improvement Plan (PIP) [3] was established in 2012. The goal is to deliver 2.25E17 protons per hour at 15 Hz by 2016. This increase in proton throughput has to be achieved by

- maintaining 85% or higher availability;
- maintaining the same residual activation in the accelerator components.

The primary users of a high-intensity proton beam will continue be to generate neutrinos at the 8 GeV Booster Neutrino beamline and the 120 GeV Neutrino at Main Injector (NuMI) beamline.

## HIGH PROTON FLUX OPERATION

Achieving > 80 kW beam power from Booster will require increasing the repetition rate and/or the beam current per cycle. The ability to increase the Booster beam power by increasing the beam charge per cycle is primarily constraint by beam losses, decreasing Booster overall efficiency and shortfall in beam parameters that are acceptable to be cleanly and efficiently transfer from Booster to Main Injector. Therefore, PIP chose the path to increase the beam throughput by increasing the number of cycles with beam. Figure 1 shows the protons delivered per day and the integrated protons delivered since 1992 up to May 2012.

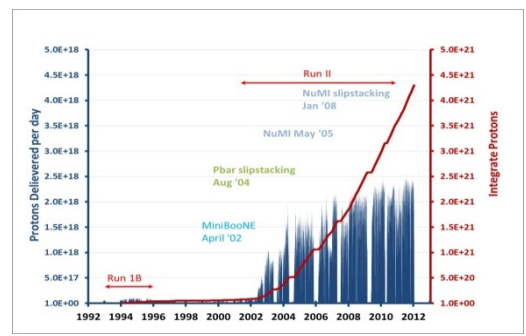

Figure 1: Booster integrated and per day protons delivery for the past 2 decades.

As can be seen, proton flux has seen a rapid increase with each yearly output exceeding the previous. Now Booster faces another challenge after the yearlong 2012 shutdown. The driving force for this long shutdown is to make the necessary Accelerator and NuMI Upgrades (ANU) for the NuMI Off-axis neutrino Appearance (NOvA) experiment [4].

___
*Work supported under DOE contract DE-AC02-76CH03000
#fgarcia@fnal.gov

NOvA is expecting up to 700 kW of peak proton power with 4.3E12 protons per batch from Booster delivered to Recycler in 12 spills at 9 Hz, where it will be slip stacked, then injected in one extraction to Main Injector for accelerating up to 120 GeV in 1.33 sec. This corresponds to 50 kW of proton power delivered by Booster. To simultaneously support the BNB physics program and the high energy neutrino program it will require to run the Booster at 15 Hz. Figure 2 shows the proton delivery projection for the next decade.

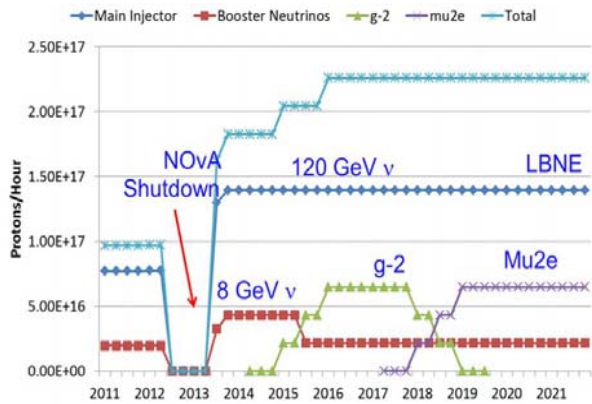

Figure 2: PIP protons economics for the next decade.

PIP will address the technical aspect that is required in order to increase the repetition rate and maintain good availability and acceptable residual activation and needs to be complete within the next five-to-six years.

The major projects that permitted Booster to run at record intensities up to now and the current limitations and going forward in order to double the proton flux are described in the remaining of this paper.

*Alignment and Aperture*

**L13 Extraction Region** Booster used to have 2 extraction regions, one at Long 13 (L13) and another one at Long 3 (L3). The former used to extract 8 GeV beam to the Booster dump, the later to the Main Injector. In both locations, the extraction scheme is a fast single-turn vertical extraction accomplished by two pairs of DC vertical bending dipoles one at each side of the septa magnet. In 2003 a feasibility study was performed in which the extraction chicane at L13 was temporarily turned off. The results were encouraging: losses were reduced by 50% leading to a better overall efficiency and increase in beam delivery. In 2006 the Booster dump was moved to 8 GeV line which allowed the removal of the L13 extraction septa and doglegs. Removal of this aperture restriction at L13 increased the beam output at the range of 15%-18% and reduced an important localized loss in the tunnel.

Outside the extraction regions, the Booster acceptance is determined by the apertures of the combined-function magnets and RF cavities and by their alignment.

**Aperture Improvement - Magnet Move** Increasing Booster aperture will decrease operational losses. Within PIP an effort is in place to, using patterns of corrector strengthens and beam loss determine the places making largest contributions to the beam orbit imperfections and formulate priorities for Booster alignment. The first attempt of using a realistic lattice model combined with latest as-found alignment data of the Booster components, the first set of magnets were realigned prior entering the current year long shutdown. The results were encouraging, with an improvement in vertical aperture at period 21 and 22 of the order of 3 mm as seen in Figure 3.

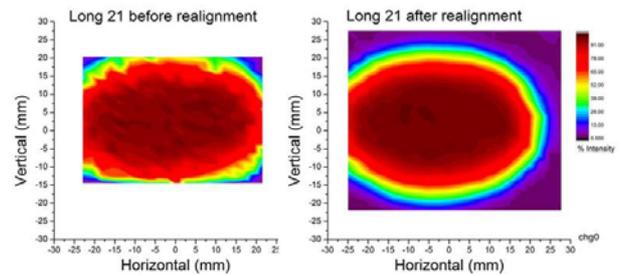

Figure 3: Aperture improvement by using lattice model combined with latest "as-found" metrology data.

*Notch and Cogging System*

Prior to 2000, beam was available in all RF buckets and during extraction, two or three bunches of beam were systematically lost in the 8 GeV extraction process as beam is swept across the magnetic septum during the extraction kicker rise time. Synchronizing a gap in the beam to the firing of the extraction kickers, seemingly simple operation is greatly complicated in Booster by the need to synchronize extraction to beam already circulating in the Main Injector. This process is known as cogging system [5] and it was essential for the success of multi-batch scheme in Main injector for slip-staking.

**Notch system** 3% of beam is removed from the buckets at low energy by using two low voltage vertical kickers, one at Long 5 and other at Long 12, that work in unison to create the notch. They fire at two different cycle times depending if the beam destination is to Main Injector or to BNB. In the last case, the gap is created at 2.4 msec and is denominated as "uncogged" event and other at 5.5 msec for "cogged" events. Therefore, notching for cogged events creates about 80% more energy loss compared to uncogged events. In both cases, the kicked beam is aimed to the collimator system. However, about 87% of the lost beam is deposited at the magnet pole tips. Booster has never lost a gradient magnet, but radiation is noticeable at this location. Going to 15 Hz operations the total beam power lost predicted in the tunnel is 270 W.

Within PIP, a two-phase approach upgrade is planned for Booster notching improvement [6]. Phase I, which is current on-going during the ANU shutdown, plans to move the notcher kickers from Long 5 to Long 12 and install a new absorber. MARS simulation has shown that by kicking the beam horizontally into the new absorber, a potential increase in notching efficiency of 10% can be obtained. During phase II, a new complement of 6 short kickers with respective new power supplies, with expectation to reach 20 nsec rise time will replace the existing system.

**Cogging system** As mentioned above, cogging allows the synchronized beam transfers to Main Injector RF bucket. The present cogging scheme is performed by control of the beam radial position via adjustments of the RF frequency. With the upgrade of new Booster corrector package [7], momentum cogging will be possible. This scheme will utilize the horizontal dipole correctors located at the short straights to change the magnetic field during the acceleration cycle, maintaining fixed the horizontal orbit in the Booster, therefore saving aperture. The positive effects one can obtained if successful are tremendous, such as notching creation at injection energy for cogged events, reducing operational losses and therefore, making notch outside the Booster tunnel possible.

*Collimation System*

In 2004 a two-stage collimation system was implemented to aid deposition of uncontrolled losses at a location that is well shielded protecting essential machine elements. The system was designed to intercept high amplitude particles by removing the halo and depositing on well shielded area reducing radiation of the rest of the machine to acceptable levels. However, this design is not compatible to the frequent radial orbit variations inherent in the RF cogging scheme. This is another potential improvement in the future to be vital for 15 Hz operation once the momentum cogging is implemented and made operational.

*Shielding assessment*

Booster Safety Assessment Document (SAD) had to be completely revised with the start of the neutrino program over a decade ago. Above ground radiation was a major concern during the planning to increase intensities, especially aggravated by the misfortune of the L3 extraction region be located below the Booster West Towers, where personnel offices are present. Therefore, extraction losses became a bigger concern and additional shielding was required on top of the extraction region and around the Booster Tower foundation.

Currently the SAD is under review to evaluate what type of measurements we need to implement in order to safely move towards 15Hz operations. One of the biggest concerns found was related with the 182 single-leg cable penetrations current filled with 12ft of polyethylene beads. According to the simulation results, the most penetrating radiation component is neutrons. At the time of this report, we haven't found a satisfactory solution.

*Beam loss control*

Radiation damage to Booster components will remain an on-going concern especially at expectations of higher proton throughput. Booster is equipped with an array of 50 interlock loss monitors arranged around the ring. Each loss monitor is carefully set to a limit observed losses at a particular region. In addition to the loss monitors, the average beam power loss is calculated by measuring the number of protons in the ring, weighted by the beam energy and integrating it through the cycle. Presently the limit is set to 525 W.

## CONCLUSIONS

The Fermilab Booster has made tremendous improvements toward meeting the demand of the Fermilab physics program. Another factor of 2 in protons deliverable is expected from Booster to support the future physics program at the laboratory. Within PIP a number of improvements have been planned and there is optimism that Proton Source will be able to reach these goals.

## ACKNOWLEDGEMENTS

This work would not have been possible without the dedication and tireless efforts of many members of the Proton Source Department, Accelerator Division Support Departments, Accelerator Physics Center (APC), Technical Division (TD) and Particle Physics Division (PPD). The author is grateful for helpful discussions with Alexandr Drozhdin, Salah Chaurize, Bill Higgins, Duane Newhart, William Pellico, Igor Rakhno, Kiyomi Seiya and Todd Sullivan.